\documentclass[
    ,final            
  ]
  {aipproc}

\layoutstyle{8x11single}
\usepackage{graphicx}
\usepackage{graphics,epsfig}
\usepackage{bm}
\usepackage{amssymb}
\usepackage{wasysym}

\begin{document}

\title{kGamma distributions in granular packs }


\author{ T. Aste}{
address={School of Physical Sciences, University of Kent, Canterbury, Kent, CT2 7NH, UK.},
altaddress={Department of  Applied Mathematics, Research School of Physical Sciences and Engineering, The Australian National University, 0200 Canberra, ACT, Australia.},
altaddress={King's College Department of Mathematics, Strand London WC2R 2LS, UK.}
}
\author{ G. W Delaney}{address={CSIRO Mathematical and Information Sciences, Private Bag 33, Clayton South, Vic, 3168, Australia.}}
\author{ T. Di Matteo}{address={King's College Department of Mathematics, Strand London WC2R 2LS, UK.}}



\begin{abstract}
It has been recently pointed out   that local volume fluctuations in granular packings follow remarkably well a shifted and rescaled Gamma distribution named the \emph{kGamma} distribution [T. Aste, T. Di Matteo, Phys. Rev. E 77 (2008) 021309].
In this paper we confirm, extend and discuss this finding by supporting it with additional experimental and simulation data. 
\end{abstract}

\maketitle

\section{Introduction}
The description and understanding of amorphous structures is very challenging because the lack of any translational symmetry makes it hard to encode the structural information into a compact form.
The absence of periodicity does not exclude however repetitiveness. 
Indeed, in amorphous systems several local configurations, or `motifs', are repeated often. 
However, one must consider that these `repetitions' typically concern similar but non identical units with small variations between one and the other. 
Furthermore, the choice of the parameters that identify the local motifs is somehow arbitrary and, as a consequence, depending on the detail of the description,  one can  gather information about the local structure at different levels.
The identification of these motifs and the study of their variations are the fundamental first step toward the understanding of the structure of granular materials and of other disordered systems such as amorphous phases. 
In these disordered systems, the structure is necessarily defined in statistical terms and it can be characterized by the probability of occurrence of a given motif.
Statistical mechanics is the theoretical instrument to calculate such probability.  

In this paper we focus on equal-sized bead packings both from experiments and simulations.
For these systems we identify the local structural motifs with the Voronoi cells which are defined as the portion of space closest to a given bead center than to any other in the packing. 
In particular, we choose to consider the volume  $V$  of such cells as the only parameter identifying the local structural organization. 
The overall packing fraction $\phi$ of the sample is directly related to the average Voronoi volume  $\left< V \right>$ by $\phi=V_s/\left< V \right>$ with $V_s = \pi d^3/6$ the volume of a bead with diameter $d$. 
The Voronoi cell defines a region of pertinence around each bead.
The fluctuations of the Voronoi volumes are therefore a measure of the local variations in the packing fraction.
Indeed, the local fraction of occupied volume is $\varphi=V_s/V$.
In this paper we discuss the distribution of such volumes showing that it is very well reproduced by a kGamma distribution \cite{AsteKGammaPRE08}.
Remarkably, this functional form is retrieved in a wide set of very different systems from idealized hard sphere packings to glass beads in water.

\section{Experiments and Numerical simulations}
\subsection{Acrylic beads in air}
The experimental data sets of bead configurations that we analyze in this paper are from the database on disordered packings \cite{Database} which contains structural data from experimental sphere packings obtained by X-ray Computed Tomography. 
Specifically our study concerns 6 samples (A-F) composed of acrylic beads prepared in air  within a cylindrical container with an inner diameter of $ 55\; mm$ and filled  to a height of $\sim 75\; mm$ \cite{AstePRE05,AsteKioloa,Aste05rev}.  
Samples A and C contain $\sim 150,000$ beads with diameter $d =1.00\;  mm$ and polydispersity within $0.05 \; mm$.
Samples B, D-F contain $\sim 35,000$ beads with diameter $d = 1.59 \; mm $ and polydispersity within $0.05 \; mm$.
The two samples at lower packing fraction (A, B) were obtained by placing a stick in the middle of the container before pouring the beads and then slowly removing the stick \cite{ppp}. 
Sample C  was prepared by gently and slowly pouring the spheres into the container.
Sample D was obtained by a faster pouring. 
In sample E, a higher packing fraction was achieved by gently tapping the container walls.  
The densest sample (F)  was obtained by a combined action of gentle tapping and compression from above (with the upper surface left unconfined at the end of the preparation). 
To reduce boundary effects, the inside of the cylinder was roughened by randomly gluing spheres to the internal surface.
The packing fraction of each of the samples is estimated at: A, $\phi \sim 0.586$; B, $\phi \sim 0.596$; C, $\phi \sim 0.619$; D, $\phi \sim 0.626$; E, $\phi \sim 0.630$; and F, $\phi \sim 0.640$.

\subsection{Glass beads in water}
Twelve other samples (FB12-24 and FB27) containing about 150,000 glass beads with diameters $0.25\;mm$ are also analysed. 
The packings were prepared in water by means of a fluidised bed technique \cite{Database,Schroder05,kAste06} within a  vertical polycarbonate tube with an inner diameter of 12.8 $mm$ and a length of 230 $mm$.
Packing fractions between 0.56 and 0.60 were obtained by using different flow rates with higher rates associated with lower packing fractions. 
After each flow pulse, the particles sediment forming a mechanically stable packing.

\subsection{Identification of the grain positions by X-ray computed tomography}
X-ray Computed Tomography (XCT) is used to calculate the coordinates of the bead centers.
This is done by applying a convolution method to the three-dimensional XCT density map efficiently implemented by use of (parallel) Fast Fourier Transform.
Furthermore a watershed method is also used to identify distinct grains.
With this technique the estimation of the position of the centre of mass of each grain can be achieved with a precision better than 0.1\% of the sphere diameter. This is well below the grain polydispersity that is estimated around 1 to 3~\% depending on the sample. 

\subsection{ Lubachevsky-Stillinger simulations}
A set of simulated packings are produced by using a modified  Lubachevsky-Stillinger (LS) algorithm \cite{Skoge06}. The simulation is an event-driven  Newtonian dynamics  in which the spheres are considered perfectly elastic without any rotational degree of freedom and with no friction. 
The simulation is performed in a cubic box with periodic boundary conditions, without gravity. 
During the simulation, the radii of the spheres are gradually increased from a very loose initial state to more densely packed configurations. In these simulations the principal control parameter is the growth rate for the sphere radii. 
Small values of growth rates will result in crystallisation.  
To avoid crystallization the growth rate should be rather large, forcing the packing into ``jammed'' non-crystalline structures where the spheres cannot be further expanded at finite pressure \cite{Torquato00,Rintoul98}. 
Simulations were performed by using the code at: http://cherrypit.princeton.edu/Packing/C++/  on  $N=10000$ spheres with initial temperature 0.1, with initial packing fraction $0.1$ and with a number of event per cycle equal to 20.
The spheres were expanded with different growth rates  between 2e-5 to 0.5, until a maximal reduced pressure of $10^{12}$ was reached \cite{Donev05a}.

\subsection{Discrete element method simulations}
We use Discrete Element Method simulation (DEM) which integrates the Newton equation of motion with both translational and rotational degrees of freedoms for elasto-frictional spheres under gravity \cite{hutzler2004,delaney2007,delaney2008-1}.
The spheres interact only when overlapping, with a normal repulsive force $ F_n = k_n \xi_n^{3/2}$ where $\xi_n = d - |{\vec{r}_i} - {\vec{r}_j}|$ is the overlap between grains of diameters $d$ with centres at ${\vec{r}_i} $ and ${\vec{r}_j}$ and $k_n$ is the stiffness parameter ($k_n = d/2Y/(3(1-P^2))$, with $Y$ the Young's modulus and $P$ the Poisson ratio) \cite{Landau1970,Makse04}. 
Tangential force under oblique loading is also considered as \mbox{$F_t  = -\textrm{min}(|k_t \xi_n^{1/2} \xi_t |, |\mu F_n |) \cdot \textrm{sign}(v_t)$},  with $ \xi_t =  \int_{t_0}^t v_t (t') \, dt'$  the displacement in the tangential direction that has taken place since the time $t_0$ when the two spheres first got in contact, where $v_t$ is the relative shear velocity and $\mu$ is the kinematic friction coefficient between the spheres and $k_t$ the tangential stiffness parameter typically assumed $2/7 k_n$ \cite{Cundall1979}.  
Normal visco-elastic dissipation   $F_n = - \gamma_n \xi_n^{1/2} \dot{\xi}_n$ (with $\dot{\xi}_n$ the normal velocity) and a viscous friction force $F_t = - \gamma_t v_t$ \cite{Schafer96} are also included. 

Here we report data for 64 simulations prepared by pouring, into a cylinder with a rough boundary, spherical beads with diameters $3$mm.
The cylinder had section $\approx 22 d$ and it was filled with $9614$ beads to an height of $\approx 56 d$ resulting in an initial packing fraction  around $0.25$. 
The beads were let sediment under gravity reaching a final mechanically stable state with packing fraction in a range between $0.55$ to $0.64$ depending mainly on the value of the friction coefficient (larger frictions smaller packing fractions), and also on the gravity and on the stokes constant.
The time step has been set at  8.0e-6 sec, the grain mass is 0.003 kg and $k_n=$1.9e7, $k_t=$5.6e6. 
The simulations were performed at various $k_s$ from 1e-3 to 1e2, various values of gravity from $1$ to $10$ and several friction coefficients from 1e-4 to 1e4.

\section{Equilibrium statistical mechanics prediction for the Voronoi volume distribution}
 \label{s.maxent}
\label{s.4}

Granular structures are disordered. 
This means that, differently form crystals,  a unique ``ideal'' structure where all the grains positions are uniquely assigned does not exist.
There are instead a very large number of structures that have equivalent global properties (packing fraction, mechanical properties, etc.) but differ in the way the grains are locally arranged.
For these disordered packings we aim to find a relation between global functional properties and local structural properties and identify the probability of occurrence of specific local structural features for given global properties.
Statistical mechanics is the theoretical framework that allows us to perform this kind of computation.

A statistical mechanics approach for granular systems was firstly proposed by Edwards in 1989 \cite{Edwards89}.
The key idea is that these non-thermal systems can be described by using a formalism very similar to the one developed for molecular gasses by substituting the constraint on the energy with a constraint on the volume occupied by the  system.
Although this is one of the few examples of extension of classical statistical mechanics concepts to a-thermal systems, the Edwards' approach is rather straightforward. Any reader with some familiarity with thermal physics and classical statistical mechanics will recognize that the forthcoming statistical description of granular systems is formally identical to the one for molecular gasses with `$E$'  substituted with `$V$'.
However, conceptually, the approach is not trivial because in granular systems we lack mechanisms equivalent to temperature and molecular chaos that  allow thermal systems to explore homogeneously the phase space.
In granular systems energy is dissipated in inelastic collision and the system soon reaches a static state with immobile grains at mechanical equilibrium.
Such state can only be changed by perturbing the system, injecting energy, for instance through vibrations or fluid flow \cite{Jaeger96,Schroder05}.
For a given preparation protocol one aims to identify the probability of occurrence of some specific structural features and their related functional properties.
In order to associate a probability to a given structure, one should (virtually) explore the whole set of possible structures which are attainable through a given preparation protocol and compute the frequency of occurrence of that specific structure within the ensemble of all attainable realizations. 
Within equilibrium statistical mechanics approach this computation is typically performed by assigning an entropy and maximizing it; finding in this way the configurations with maximum likelihood. 
It is beyond the propose and the possibility of this paper to fully expose the subdue issues around this kind of approach that have been debated  in the literature for the last twenty years since 1989 \cite{Edwards89,Mehta89,Edwards94,Edwards99,Brujic03,Blumenfeld03}. 
Here we are merely comparing the theoretical prediction from a statistical mechanics approach (namely Eq.\ref{e.pV1D}) with data from experiments and computer simulations.
To have a better insight of our approach to this problem and for a formal deductive derivation of Eq.\ref{e.pV1D}, the interest reader can refer to  \cite{AsteDeductiveSM} and references therein.
Let us hereafter only briefly sketch the main ideas and the main passages  of this approach.

In analogy with the Edwards' original approach here we consider the ensemble of mechanically stable configurations that can be achieved by means of a given preparation protocol resulting in a given average packing fraction over a large number of trials \cite{Edwards89,Mehta89,Edwards94,Edwards99,Brujic03,Blumenfeld03}.
Here we look at the statistics of the local configurations of each Voronoi cell and we maximize entropy to calculate the probability $p(V)$ for a cell with volume $V$ in a sample with packing fraction $\phi$ where the average Voronoi  volume is $\left< V \right> = V_s/\phi$.  
A classical equilibrium statistical mechanics theory `a la Edwards' gives:
\begin{equation}
p(V) = \frac {  \Omega(V) e^{-   \beta V   }   }{  \sum_{V'} \Omega(V') e^{- \beta V '  } }\;\;\;,
\label{e.ME}
\end{equation}
with $ \beta $ a Lagrange multiplier which is determined  by the constraint on the average volume: 
\begin{equation} \label{e.VmeL}
\left< V \right> =  \sum_{V} V p(V)  \;\;.
\end{equation}
Here the challenge is to compute $ \Omega(V) $ which is the `density of states' counting the number of microscopic configurations associated with a Voronoi volume $V$. 

Let us note that Eq.\ref{e.ME} is the analogous for these non-thermal systems to the Boltzmann distribution for molecular gasses where in this case the particle energy is substituted with the Voronoi cell volume.  

\begin{figure*} 
\centering
\begin{tabular}{c}
\includegraphics[width=0.95\textwidth]{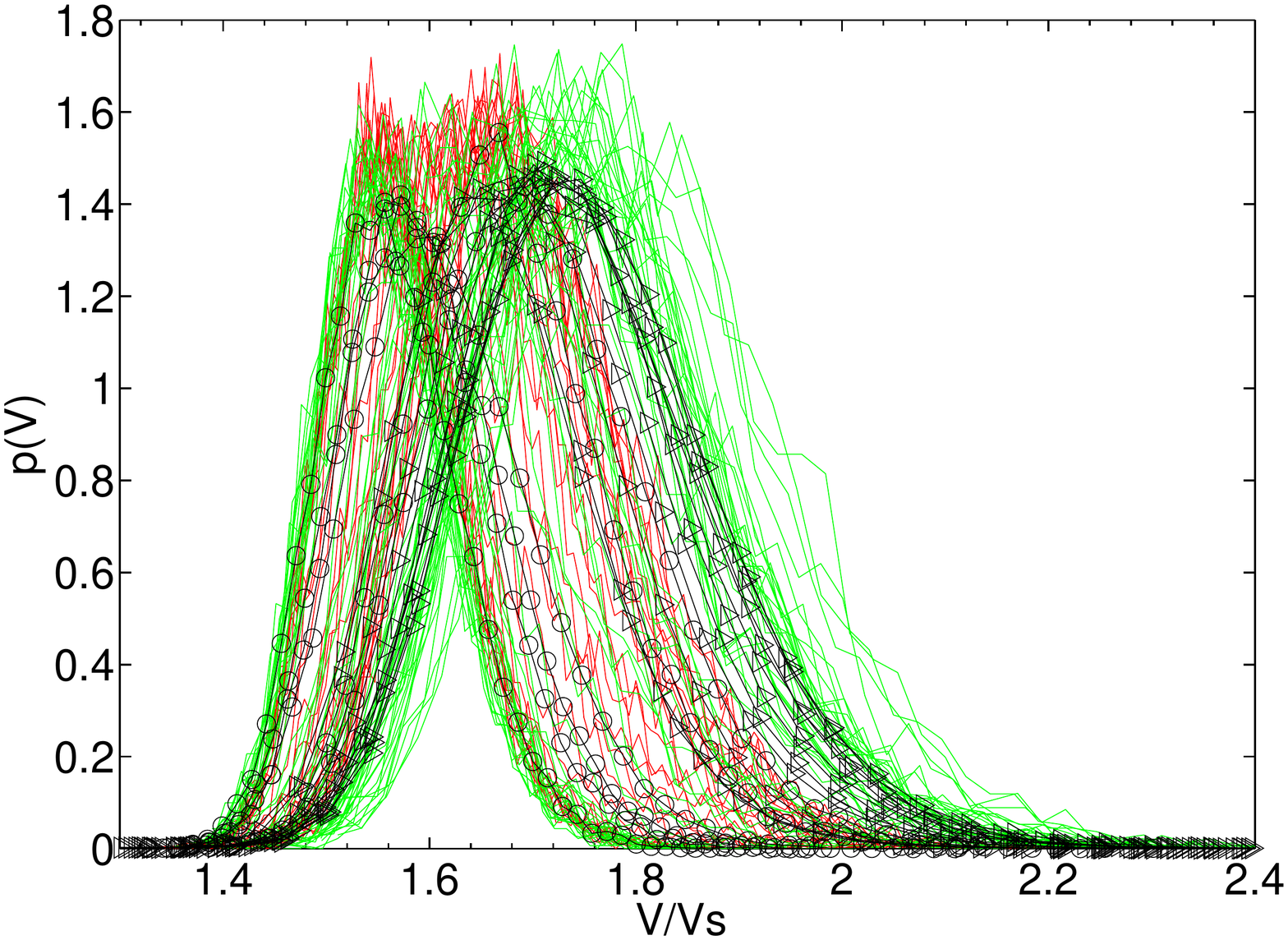} \\
\includegraphics[width=0.95\textwidth]{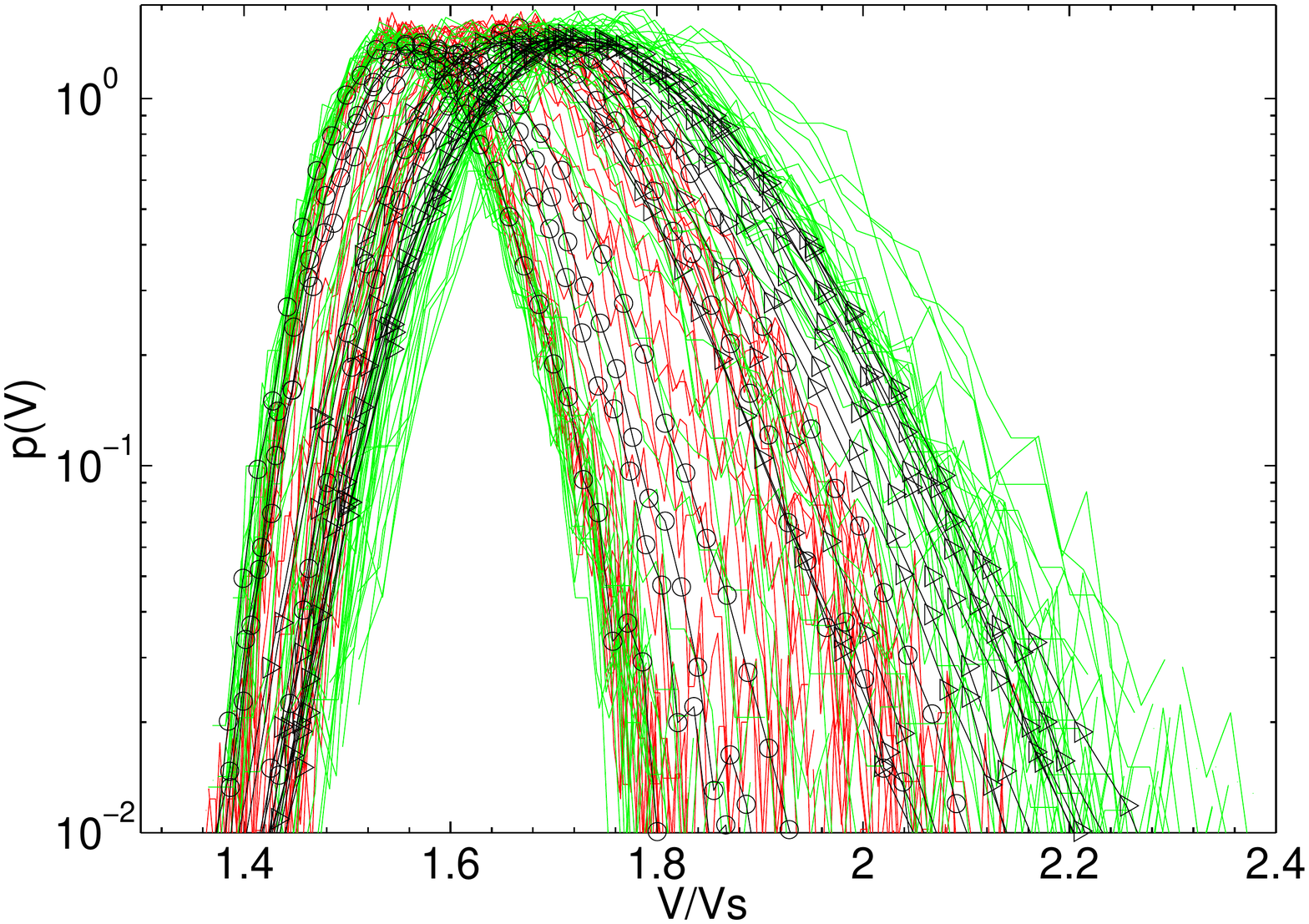}
\end{tabular}
\caption{\footnotesize (Color online).  
(top) Vorono\"{\i}  volume distributions from all the experimental and simulation data. 
The $\circ$ refer to dry acrylic beads experiments and the $\triangleright$ refer to glass beads in water.
The read lines (color online) are Lubachevsky-Stillinger simulations and the green lines are DEM simulations.
$V_s=\pi/6d^3$ is the volume of a spherical bead.
(bottom) Same plot in semi-log Y scale.}
\label{f.LSvor}
\end{figure*}

\begin{figure*} 
\centering
\begin{tabular}{c}
\includegraphics[width=0.95\textwidth]{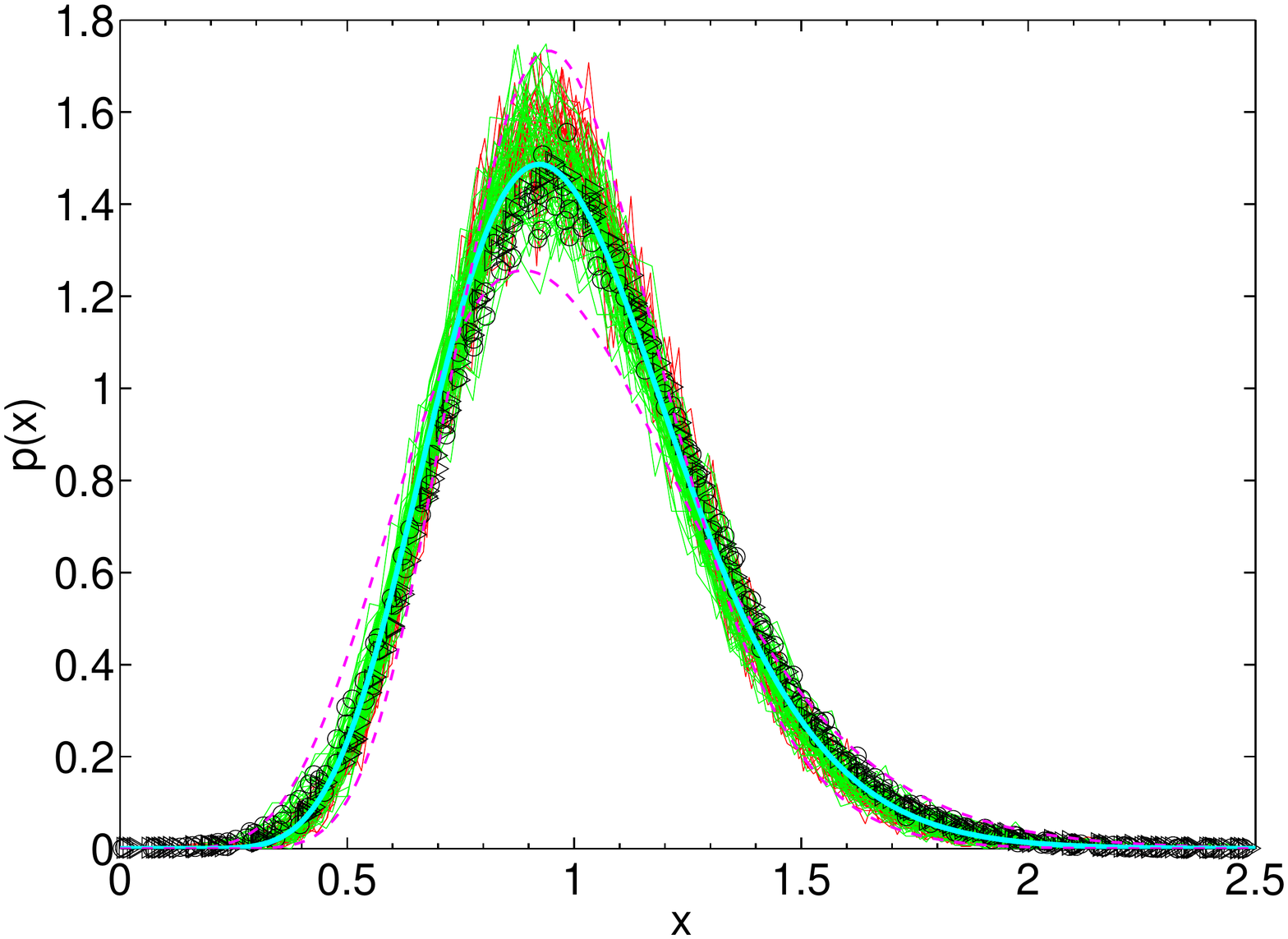} \\
\includegraphics[width=0.95\textwidth]{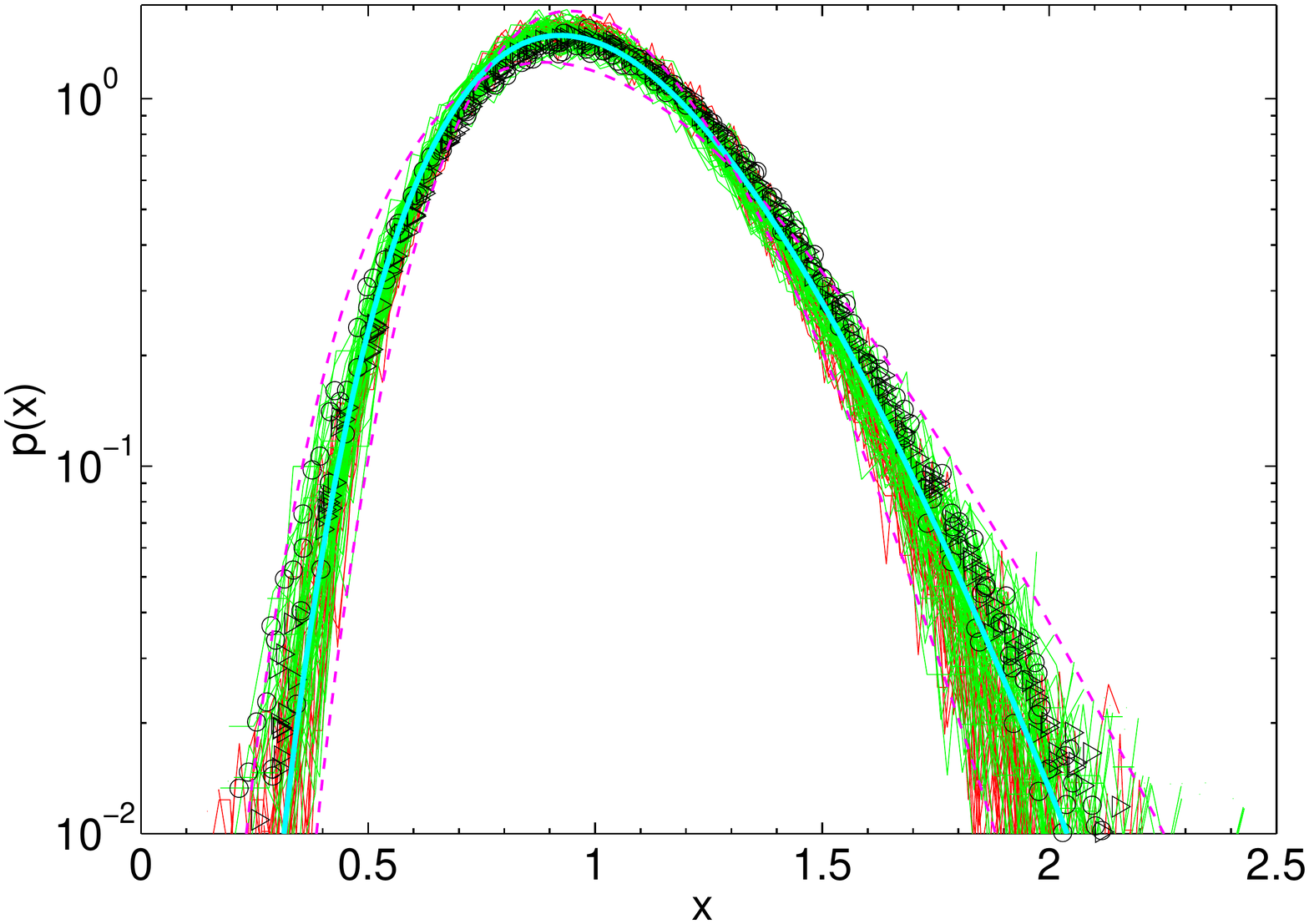}
\end{tabular}
\caption{\footnotesize (Color online).  
(top) Same Vorono\"{\i} volume distributions as in Fig.\ref{f.LSvor} but plotted vs. $x = (V - V_{min})/( \left< V \right> - V_{min})$.
(bottom) Same plot in semi-log Y scale.
The tick full line is the kGamma function $p(x)= k^k/\Gamma(k) x^{k-1} \exp(-kx)$ for $k=13$.
The two dashed lines are two kGamma functions for $k=9$ and $k=18$.
}.
\label{f.LSvorColl}
\end{figure*}

\subsection{Explicit derivation of the probability distribution from a simple hypothesis}
\label{s.5}

In order to compute $ \Omega(V) $  here we use a very simple hypothesis: there are $k$ `degrees of freedom' contributing to the volume of each Voronoi cell.
The idea is that each Voronoi cell is composed of $k$ elements each one contributing independently to the cell volume $V$.
Each of these `elementary cells' can have arbitrary volumes larger than $v_{min}$ under the condition that their combination must add to a total volume $V$.
Under this assumption $\Omega(V)$  can be computed exactly:
\begin{equation}
\Omega(V) = \frac{1}{\Lambda^{3k}} \int_{v_{min}}^V dv_1 \int_{v_{min}}^V dv_2 .... 
 \int_{v_{min}}^V dv_k \delta(v_1+v_2+...+v_k - V) 
 =   \frac{(V- k v_{min} )^{k-1} }{ \Lambda^{3k} (k-1)!} \;\;\;,
\label{e.S1Dbb}
\end{equation}
with $\Lambda$ a constant analogous to the Debye length.
Substituting into Eq.\ref{e.ME}, and by using Eq.\ref{e.VmeL} we obtain for the Lagrange multiplier $\beta =k/( \left< V \right> - k v_{min})$ and  the probability $p(V)$ takes the form:
\begin{equation}
p(V)  = 
 \frac{k^{k}}{\Gamma(k) } \frac{(V - V_{min})^{(k-1)} }{( \left< V \right> - V_{min})^{k}} \exp \left( {-k \frac{V-V_{min}}{ \left< V \right> - V_{min} } } \right)\;\;\;,
\label{e.pV1D}
\end{equation}
with $V_{min} = k v_{min}$ which is the minimum volume attainable for a Voronoi cell in the packing.
For a packing of equal spheres $V_{min}$ is exactly known being the volume of a dodecahedral cell circumscribing the sphere which is  $V_{min}= 5^{(5/4)}/ \sqrt{2(29+ 13 \sqrt{5}) } d^3 \simeq 0.694 d^3$ \cite{ppp}.   
Eq.\ref{e.pV1D} is a Gamma distribution in the variable $V- V_{min}$; it is characterized by a `shape'  parameter  $k$ and a `scale'  parameter $( \left< V \right> - V_{min})/k$ \cite{Gamma}. 
We call such a function: \emph{kGamma} distribution \cite{AsteKGammaPRE08}.
Interestingly, a mathematical study for the Voronoi statistics in two dimensional point processes predicts a gamma distribution for the cell area distribution \cite{Brakke}.

The mean of the distribution $p(V)$ is  $ \left< V \right>$ and its variance is
\begin{equation}
\sigma^2_v = \frac{( \left< V \right>-V_{min})^2}{k} \;\;\;,
\label{e.variance}
\end{equation}
which implies
\begin{equation}
k =  \frac{( \left< V \right>  -V_{min})^2}{\sigma^2_v}\;.
\label{e.k}
\end{equation}
This last equation gives directly the parameter $k$ from a measure of the variance of the distribution.
Therefore, there are no free fit parameters in Eq.\ref{e.pV1D}.

It might be of some use to compute explicitly the related distribution for the local packing fraction $\varphi$ which is given by the identity $p(\varphi)d\varphi = p(V)dV$ yielding to 
 \begin{equation}
p(\varphi)  = 
 \frac{k^{k}}{\Gamma(k) } \frac{\varphi_{max}}{\varphi^2} \frac{(\varphi_{max}/\varphi - 1)^{(k-1)} }{( \varphi_{max}/\phi - 1)^{k}} \exp \left( {-k \frac{\varphi_{max}/\varphi - 1}{ \varphi_{max}/\phi - 1} } \right)\;\;\;,
\label{e.pV1Dphi}
\end{equation}
with $\varphi_{max}=V_s/V_{min}\simeq 0.75$ the maximum attainable local packing fraction in equal sphere packings.

\begin{figure*} 
\centering
\begin{tabular}{cc}
\includegraphics[width=0.55\textwidth]{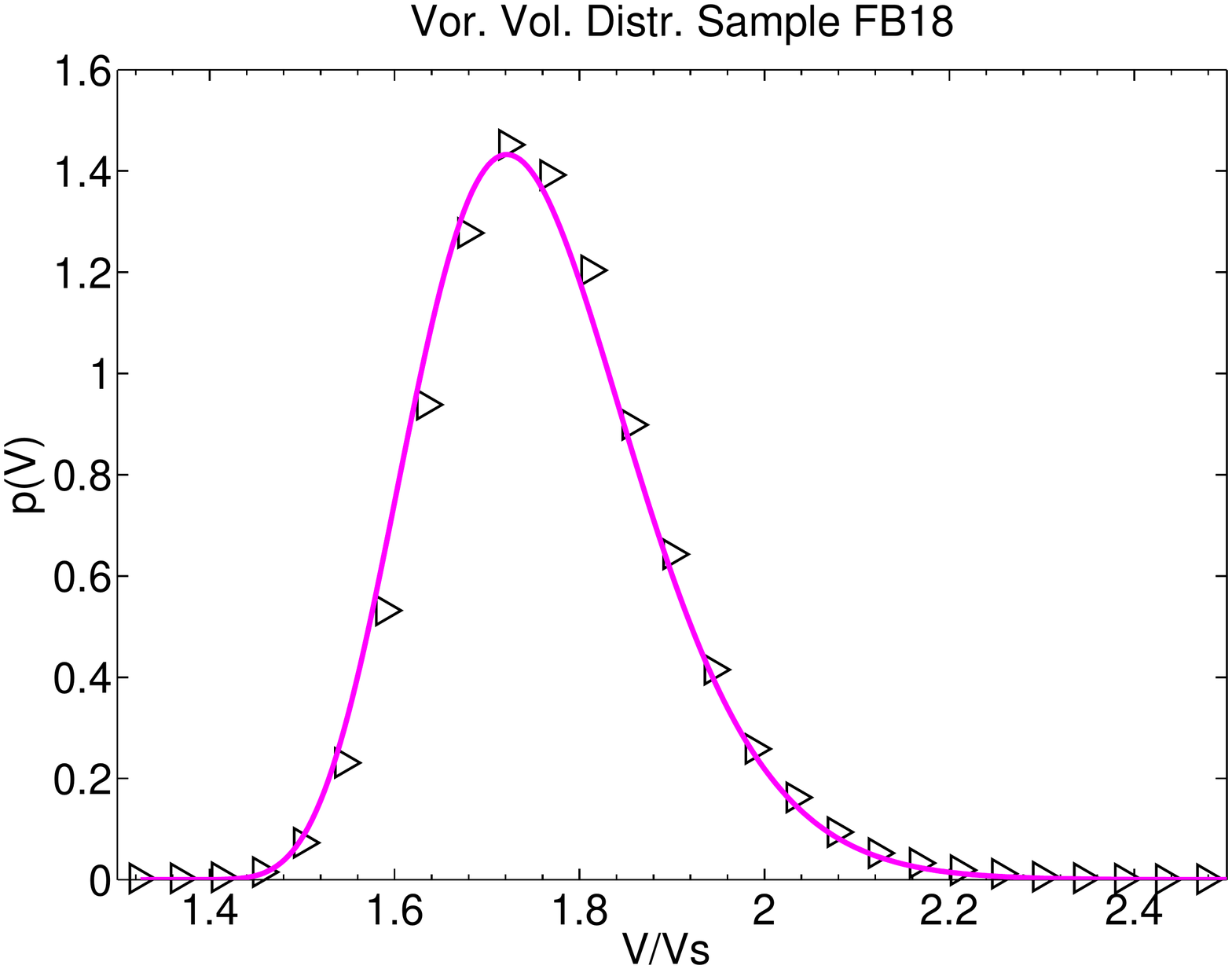} &  \includegraphics[width=0.55\textwidth]{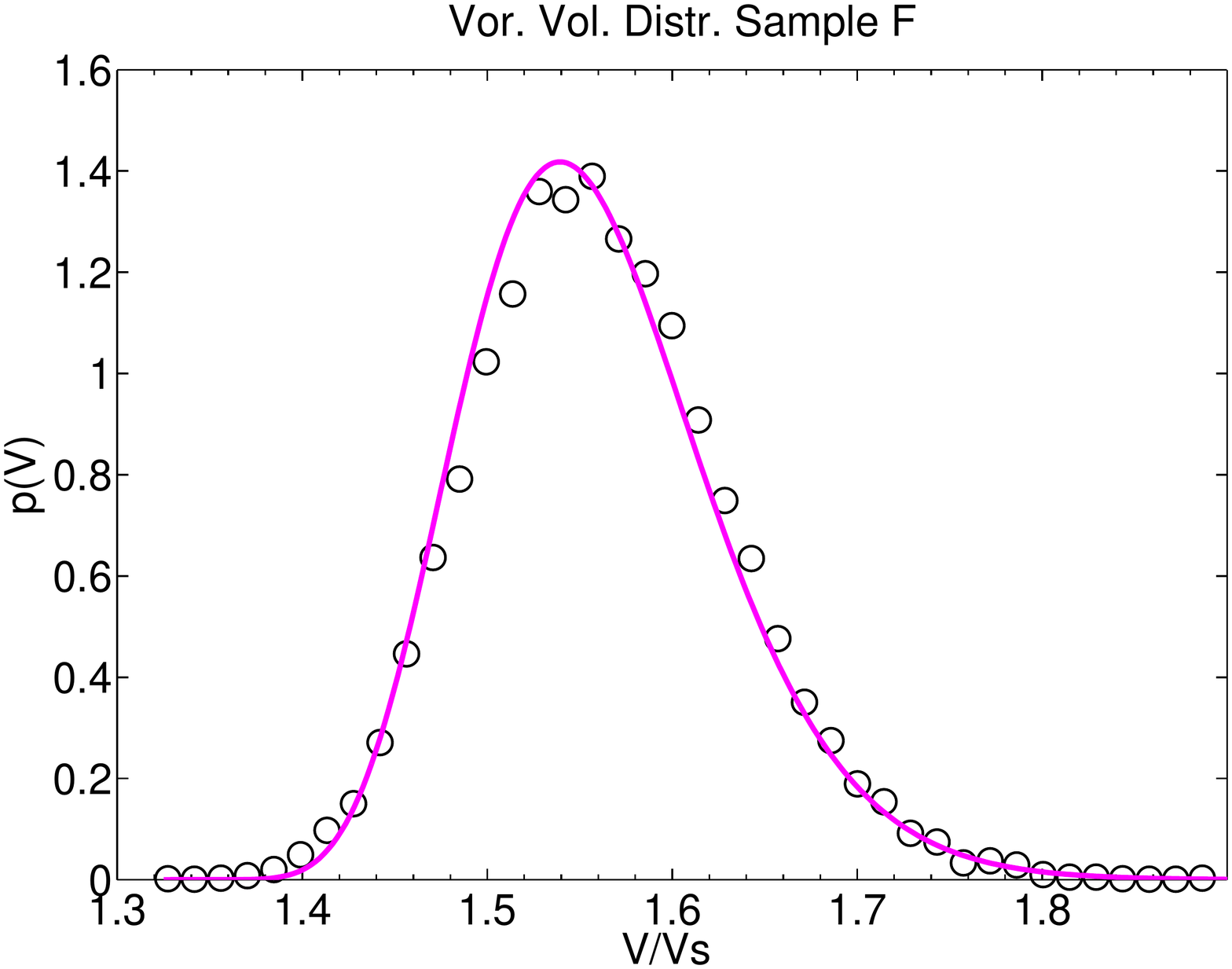} \\
\includegraphics[width=0.55\textwidth]{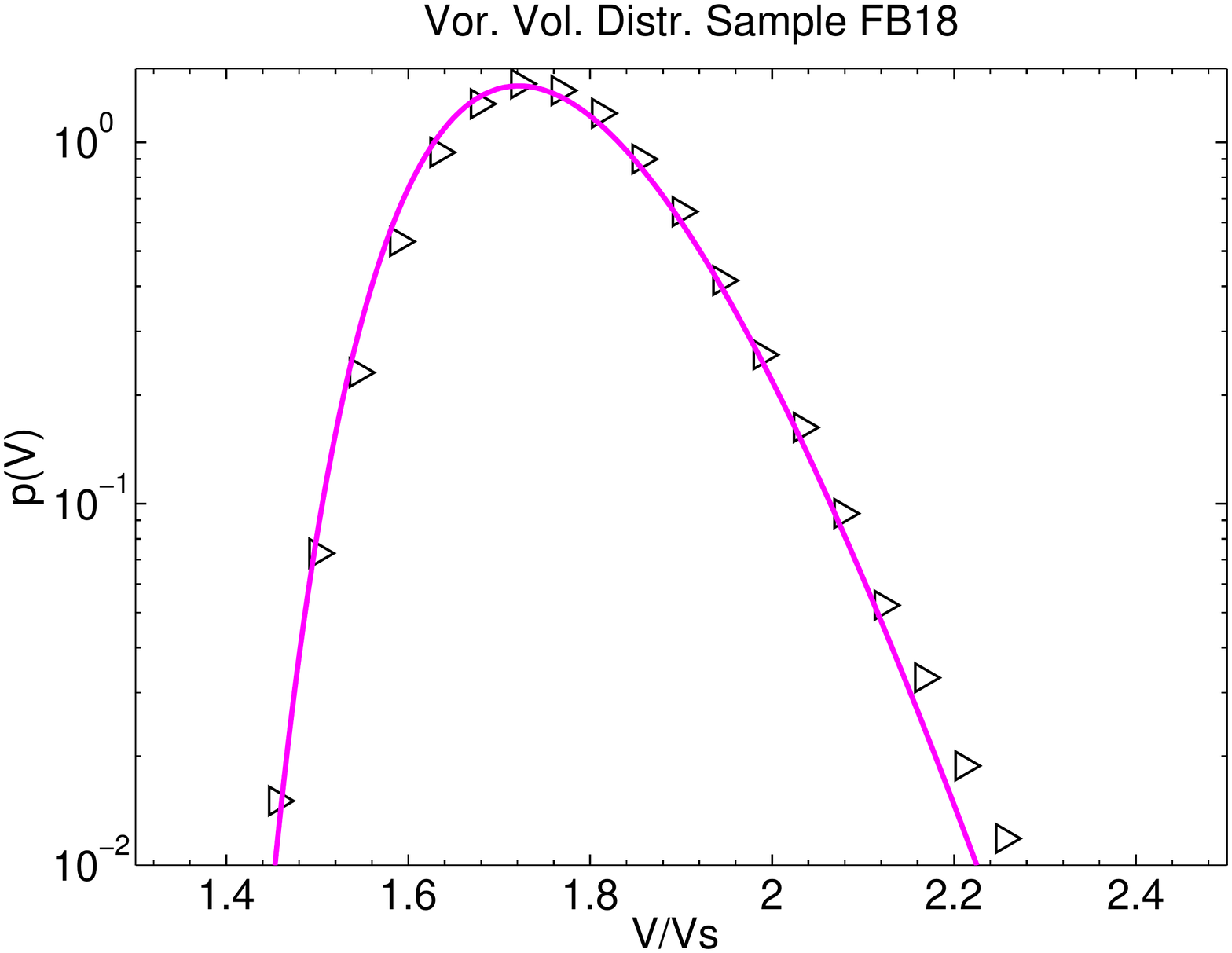} & \includegraphics[width=0.55\textwidth]{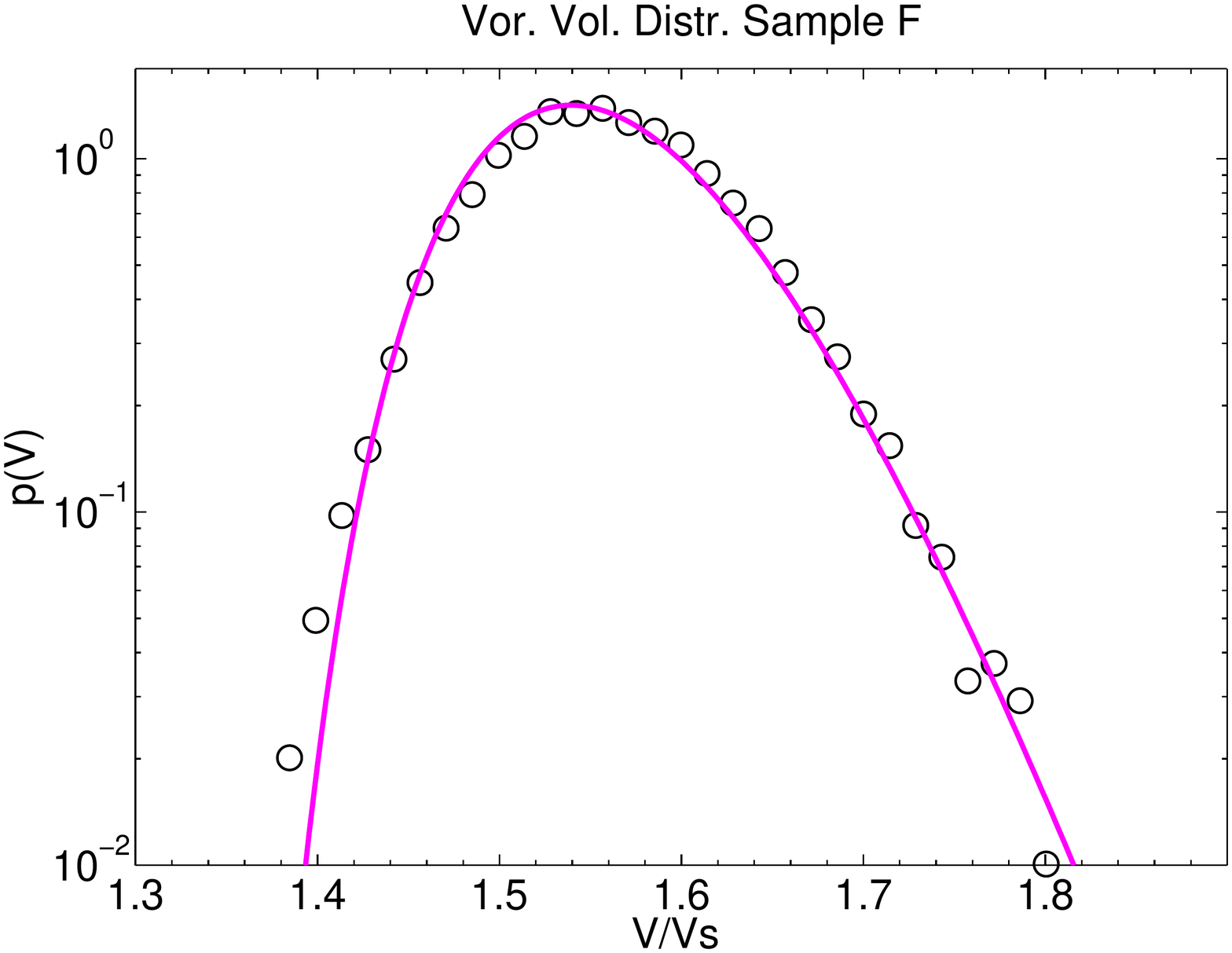} 
\end{tabular}
\caption{\footnotesize (Color online).  
(left) Experimental Vorono\"{\i}  volumes distributions  for the sample with lowest packing fraction obtained by using fluidized bead technique for glass beads in water (FB18). 
(right) Experimental Vorono\"{\i}  volumes distributions  for the sample with highest packing fraction obtained by pouring acrylic beads in air (F). 
(bottom) Same plots in semi-log Y scale. 
The lines are the kGamma distributions with $k= {( \left< V \right>-V_{min})^2}/{\sigma^2_v }$ with $\sigma_v$ the measured standard deviation and $\left< V \right>=V_s/\phi$ with $\phi$ the measured packing fraction.
There are no adjustable parameters or fits.
}
\label{f.LSvorCollExp}
\end{figure*}

\section{Voronoi  volume distributions from experiments and simulations}
\label{s.6} \label{s.7}

We have tested the validity and resilience of the kGamma behavior over a set of several hundreds numerical simulations and over 18 different experiments. 
The simulations consisted of both Lubachevsky-Stillinger newtonian dynamics of frictionless hard spheres and DEM simulations of elasto-frictional spheres.
The experiments include dry and fluidized bead samples.

Figure \ref{f.LSvor} shows the resulting distribution of the Vorono\"{\i} volumes. 
One can see that such distributions span a very broad interval of volumes with $V$ between $\approx 1.3 V_s$ and $\approx 2.5 V_s$ with large differences between different samples shown both in the average values and in the distribution spreading. 
We observe that all distributions show some degree of asymmetry around the maximum  with larger probabilities for large volume fluctuations.

From Eq.~\ref{e.pV1D}, one can see that \emph{kGamma} distributions characterized by similar values of $k$ must result into similar behaviors when plotted as  function of $x=(V-V_{min})/( \left< V \right> - V_{min} )$.
Figure \ref{f.LSvorColl} shows the plot of all the distributions as a function of such shifted-rescaled variable.
We note that all distributions  collapse into a very similar functional form which is very well described by the kGamma function $p(x)= k^k/\Gamma(k) x^{k-1} \exp(-kx)$ with $k$ ranging in a narrow interval.

\begin{figure*} 
\centering
{\includegraphics[width=0.95\textwidth]{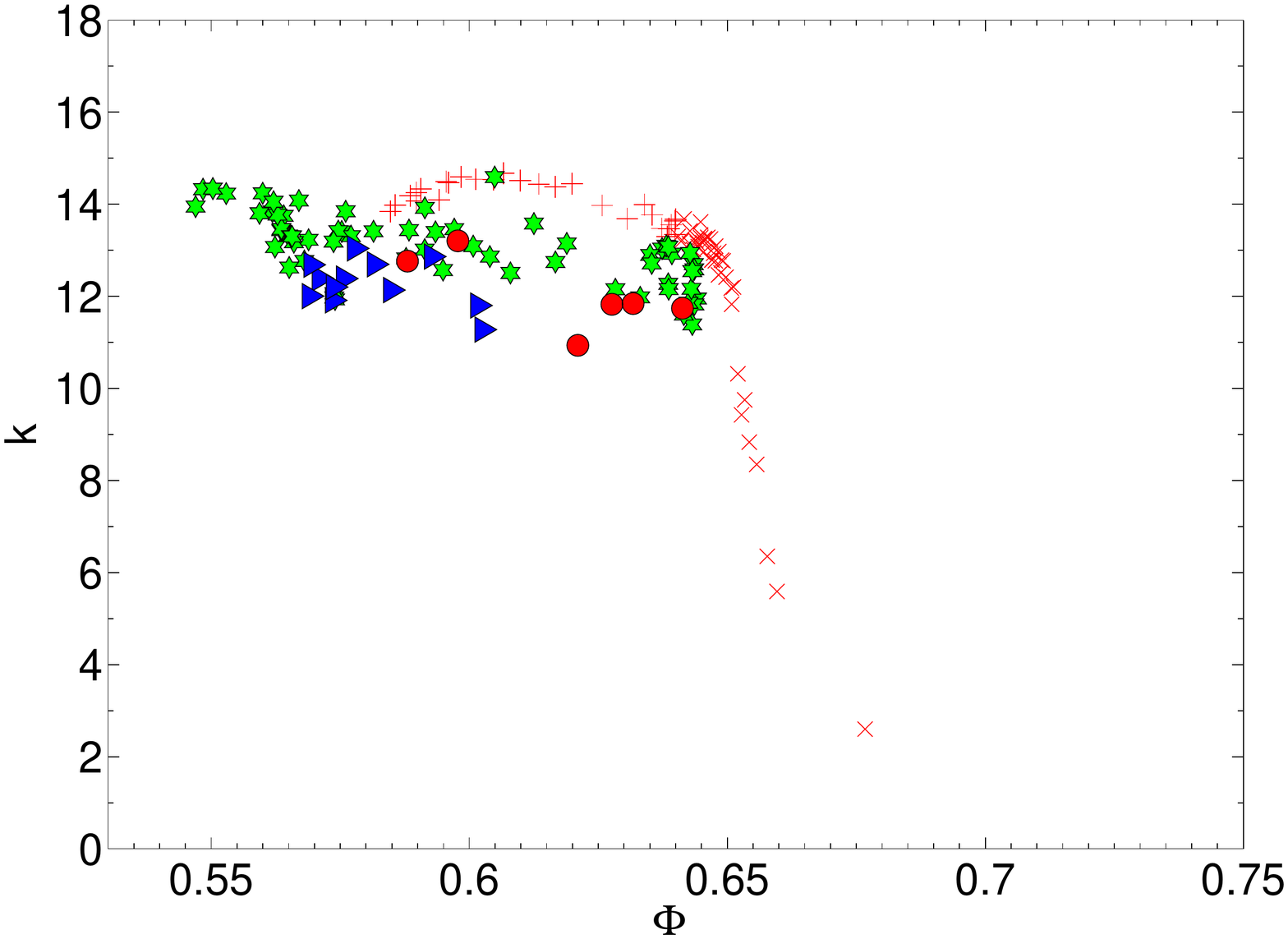}}
\caption{\footnotesize (Color online).  
Behavior of $k$ calculated from $k =  ( \left< V \right>  -V_{min})^2/\sigma^2_v$ (Eq.\ref{e.k}) vs. packing fraction $\phi$.
The $+$ refer to  Lubachevsky-Stillinger simulations, the $\star$ refer to DEM simulations, the $\circ$ refer to dry acrylic beads experiments and the $\triangleright$ refer to glass beads in water.
The $\times$ refer instead to Lubachevsky-Stillinger simulations above $\phi \simeq 0.64$ where a poly-crystaline phase starts to emerge.}
\label{f.krho}
\end{figure*}

The goodness of the description of these distributions by means of the kGamma function in Eq.\ref{e.pV1D} can be judged from Fig.\ref{f.LSvorCollExp} which shows the agreement between the prediction from  the kGamma function $p(V)$ and the measured data from experiments.  
Similar agreements are found across all samples from both experiments and simulations.
It should be stressed that in this plot there are no adjustable parameters or fitting constants.
Indeed, the only two parameters in Eq.\ref{e.pV1D} are $\left< V \right>$   and $k$  which are uniquely determined respectively from the sample packing fraction ($\left< V \right> = V_s/\phi$) and from the measured standard deviation of the Vorornoi volumes ($k= (\left< V \right>  -V_{min})^2/\sigma^2_v$, Eq.\ref{e.k}).

\begin{figure*} 
\centering
{\includegraphics[width=0.95\textwidth]{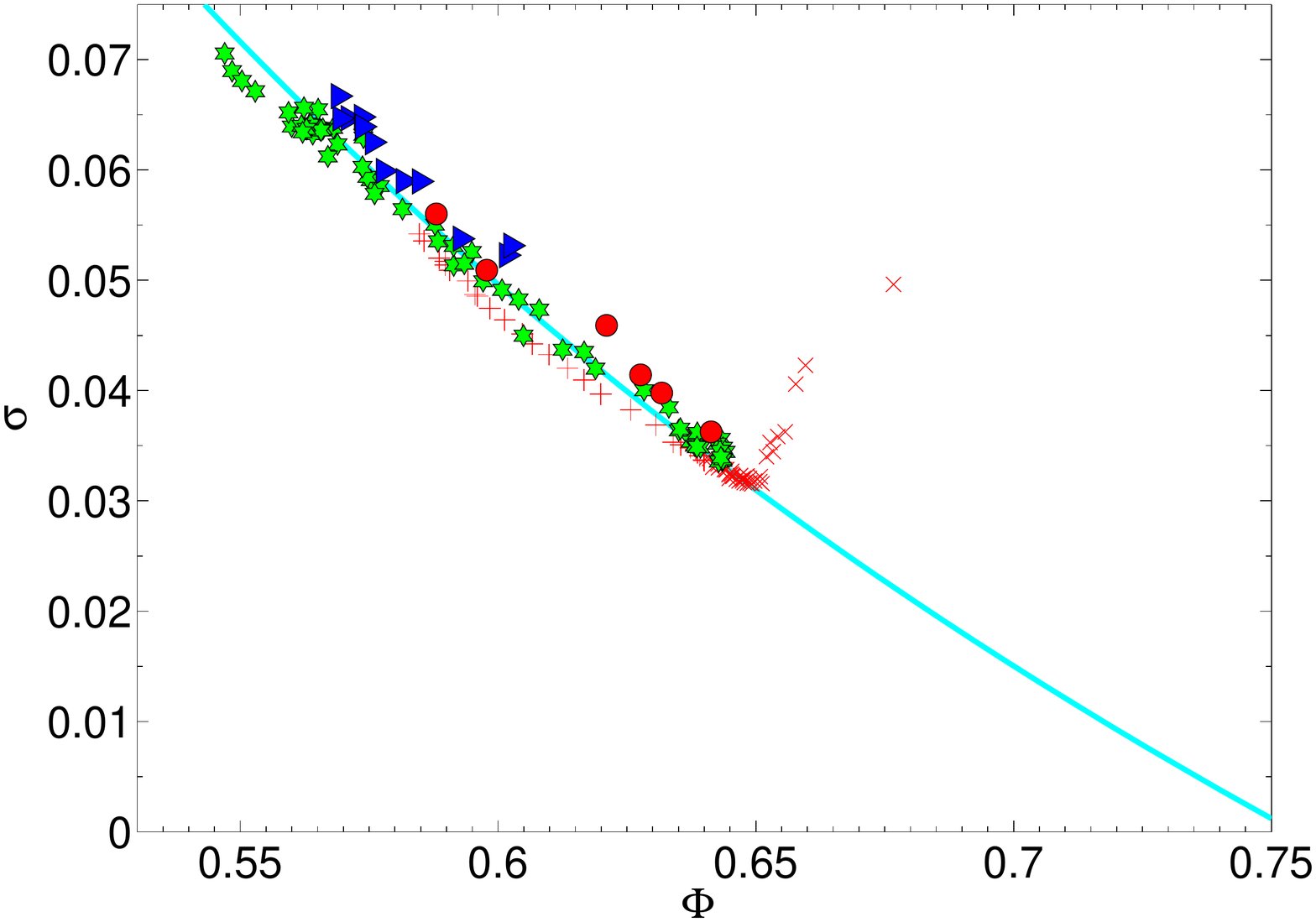}}
\caption{\footnotesize (Color online).  
Behavior of $\sigma_v$ vs. packing fraction $\phi$.
The line is the prediction from Eq.\ref{e.variance} (i.e. $\sigma = (V_s/\phi - v0)/\sqrt{k}$) for $k=13$.
The symbols are the same as in Fig.\ref{f.krho}.}
\label{f.sig}
\end{figure*}

The impressive fact of such an agreement  is that these systems are very different (ideal Newtonian spheres, elasto-frictional spheres under gravity, real experimental acrylic beads in air and glass beads in water) and they are prepared in very different ways (pouring, tapping, fluid flows, molecular dynamics, shearing).
The collapse of all these distributions around a unique functional form suggests that the main driving mechanism which determines these fluctuations is an exchange of volume among Voronoi cells that possesses only a small number of degrees of freedom. 
Accordingly with our hypothesis, such a number is given by the parameter $k$.

In Fig.\ref{f.krho} we report the measured values of $k$ as function of the sample packing fraction calculated from the relation $k =  ( \left< V \right>  -V_{min})^2/\sigma^2_v$ (Eq.\ref{e.k}). 
We observe a rather narrow and homogeneous range of values laying between 11 and 15 for all samples.
A sharp drop in $k$ is observed above $\phi \simeq 0.64$ where in the Lubachevsky-Stillinger simulations a crystalline phase starts to nucleate and grow.
Let us note that in this phase a peak at larger packing fractions appears in the volume distributions (not reported in Figs.\ref{f.LSvor} and \ref{f.LSvorColl} to avoid confusion.).

The good predictive potential of kGamma distributions for volume fluctuations in granular assemblies can also be inferred from the simple measure of the standard deviation. 
Indeed, standard deviations can be easily measured and require smaller statistical sets with respect to the whole distribution.  
Figure~\ref{f.sig} reports the trend of the measured standard deviation versus the packing fraction for all the samples.
It is clearly evident from the figure that they all follow a common decreasing trend before the crystallization onset above $\phi \simeq 0.64$. 
The line in the figure is the the prediction from Eq.\ref{e.variance}: $\sigma = (V_s/\phi - V_{min})/\sqrt{k}$), for $k=13$.

\section{Conclusions}
\label{s.9}
In this paper we have shown that kGamma distributions describe very well the observed Voronoi volume distributions in a wide variety of systems from simulated hard spheres to real experimental packings. 
The agreement  between the predicted distribution and the measures is remarkable also considering that there are no adjustable parameters or fitting constants.

We have briefly explained as a statistical mechanics approach can be used to retrieve such a distribution from a very simple hypothesis that the Voronoi cells in the systems have $k$ degrees of freedom associated to their volumes.
Despite the very good agreement with the empirical data that has been shown in this paper, this hypothesis is certainly quite strong and unrealistic. 
Indeed, it might be true that in average there are $k$ degrees of freedom per Voronoi cell but it is unlikely that \emph{each} Vorornoi cell has \emph{exactly} $k$ degrees of freedom.
A relaxation of this hypothesis on a more realistic ground would produce different outcomes predicting for instance other kinds of Gamma distributions.
However, the beauty and the strength of the  present approach is that we have only one parameter that is fully determined by the measured standard deviation. 
In our view, the exceptional agreement of all the experimental and numerical data with the prediction from kGamma distribution does not justify, at the present, the introduction of a more complicated theoretical framework with the consequent incorporation of extra adjustable parameters.

Future studies will focus on the effect of polydispersity and shapes \cite{delaney2008}.

\subsection*{Acknowledgements}

We thank  Antonio Coniglio, Mario Nicodemi, Massimo Pica Ciamarra, Matthias Schr\"oter and  Harry Swinney for helpful discussions.
Many thanks to T.J. Senden, M. Saadatfar, A. Sakellariou, A. Sheppard, A. Limaye for the help with tomographic data.

\bibliographystyle{aipproc}

\end{document}